# Empirical Analysis of the Impact of Legal Tender Digital Currency on Monetary Policy -Based on China's Data

Tiantian Zhao

**Abstract**

This paper takes the development of China's Central bank digital currencies as a perspective, theoretically analyses the impact mechanism of the issuance and circulation of Central bank digital currencies on China's monetary policy and various variables of the money multiplier; at the same time, it selects the quarterly data from 2010 to 2022, and examines the impact of the Central bank digital currencies on the money supply multiplier through the establishment of the VECM model. The research results show that: the issuance of China's Central bank digital currencies will have an impact on the effectiveness of monetary policy and intermediary indicators; and have a certain positive impact on the narrow money multiplier and broad money multiplier. Based on theoretical analyses and empirical tests, this paper proposes that China should explore a more effective monetary policy in the context of Central bank digital currencies in the future on the premise of steadily promoting the development of Central bank digital currencies.

# 1 Introducere

The introduction of legal tender digital money to accommodate the quick development of the digital economic age trend of the times is a significant invention in human history. This is made possible by the constant development and use of blockchain, 5G, and other information technologies.In the scenario, the Central Bank digital currencies with the combination of the Central Bank digital currencies and everyone's life will be higher and higher by January 4, 2022, when China's digital RMB APP is on the shelves and the Central Bank digital currencies have truly begun to enter the public's daily consumption activities. TheCentral Bank digital currencieswill, however, make it more difficult for the central bank to set reasonable money supply targets, which in turn affects how well monetary policy targets are implemented. At the moment, China primarily uses the money supply and interest rate as the intermediary index of monetary policy transmission. In order to enhance the effectiveness of China's monetary policy in controlling the macroeconomy and fostering the growth of its digital economy and inclusive finance, it is therefore of great theoretical and practical significance to study the impact of Central Bank digital currencies on monetary policy and the money multiplier.This paper may exist in the innovation: one is the innovation of research perspective, the existing literature on legal tender is mainly focused on the connotation of legal tender, regulation, and the impact on the macroeconomy, while the Central bank digital currencies on monetary policy and the impact of the money multiplier research empirical analysis is less. Secondly, in terms of empirical analysis, this paper uses the economics of elasticity and differential analysis of the impact of the Central bank digital currencies on the money multiplier of each variable to theoretically prove the establishment of the VECM model, empirically test the impact of China's legal tender on the money supply multiplier effect, and explore the long- and short-term impact of legal tender on the money multiplier.

# 2 Review of literature

Research on topics related to central bank digital currencies has gained attention in the theoretical community due to the rapid development of the digital economy and technology. New studies and literature are constantly being published that address the definition of central bank digital currencies, their issuance methods, and their effects on monetary policy.

Reviewing the development history of the Central bank digital currencies, scholars have respectively clarified the definition of the Central bank digital currencies from the perspectives of the issuer of the Central bank digital currencies, specific forms, etc. Koning (2017)[8]proposes to categorize the Central bank digital currencies into two manifestations according to whether it is based on a central bank account or not: the Central bank digital currencies (CBDC) and central bank digital account (CBDA). Bordo &Levin (2017) [3]argue that CBDB is a cost-free medium of exchange with a safe-storage value and a stable unit of account. Fan Yifei (2016)[7] argues that the Central bank digital currencies is essentially a pure credit currency that maintains the basic attributes and main features of cash notes. Yao Qian (2016)[15] points out that the Central bank digital currencies are issued by a country's central bank and circulate at the same time as banknotes with the advantages of flexible and convenient transactions and payment-as-settlement, which belongs to the category of cash in circulation. Barrdear & Kumhof (2016)[1] define digital currencies as any form of electronic money characterized by a distributed ledger and a decentralized payment system or an exchange medium. The Bank for International Settlements (2018)[2] released a report on "Central Bank Digital Currency", which puts forward the concept of the "flower of money", which is elaborated from four perspectives, namely, the issuing body is the central bank or other financial institutions, the form of the currency is digitized or physical, whether it is widely available, and whether the technological means is based on the account or based on the tokens. It is proposed that the legal tender digital currency is the center of the "flower of money". Wu Xinhong et al.(2022)[12] argue that legal tender is the Central bank digital currencies issued by the central bank based on cryptography and special transaction algorithms, endorsed by the state's credit, and has the same legal reimbursement as the national credit banknotes.

Current research considers that the Central bank digital currencies are mainly issued in one-tier and two-tier issuance modes. Fan Yifei (2016)[7] argues that there are two main issuance modes for the Central bank digital currencies: one is the direct issuance to the public by the central bank, the second is the traditional indirect issuance mode of "central bank-commercial bank-social public", and the two-tiered issuance mode is more suitable for the practice of China's financial system. Wu Zhifeng (2016)[13] puts forward three possible issuance modes of the Central bank digital currencies, the first mode is issued directly to the public by the central bank of a country, this mode has the advantage of reducing transaction costs, helping to ensure the central bank's ability to regulate currency issuance and avoiding the problem of "over issuance of currency", but it may disrupt the existing However, it may disturb the existing mechanism of money creation, regulation and supervision. The second mode is the cooperation between the central bank and commercial banks, the advantage of this mode is that it will not subvert the existing currency circulation system, but may weaken the central bank's control over the monetary system. The third model is for individual commercial banks to compete with each other, the advantage of which is that it can largely promote the competition of digital currencies, but it may lead to the existence of multiple forms of digital currencies in the economy and society, resulting in monetary chaos.

Scholars analyze the impact of legal tender on monetary policy from different angles. One is to study from the perspective of the overall impact of the Central bank digital currencies on China's monetary policy framework. Du Yongshan (2022)[5] and others analyze the impact of the Central bank digital currencies on the monetary policy framework from the perspective of the monetary policy transmission mechanism and argue that legal tender will improve the effectiveness of monetary policy. Fang Xiancang et al.(2020)[6]pointed out that legal tender affects the mediating indicators of quantity-based and price-based monetary policy, which leads to the conclusion that the Central bank digital currencies reduces the effectiveness of quantity-based monetary policy and enhances the effectiveness of price-based monetary policy. A part of scholars focus on studying the impact of l the Central bank digital currencies on various variables of quantity-based monetary policy. Blakstad & Robert (2018) [4]state that the Central bank digital currencies can reduce economic transaction costs, improve the efficiency of currency use, accelerate currency circulation, and increase the number of bank deposits and loans. Stiglitz (2017)[11] argues that the central bank can monitor the flow of the Central bank digital currencies and application scenarios, enhance the controllability and measurability of the money supply, form an anticipatory guideline for bank lending rates and fund flows, enhance the effectiveness and stability of the credit transmission mechanism of commercial banks, and improve the systematicity and transparency of monetary policy.Lin Chun (2019)[9] argues that the issuance of the Central bank digital currencies will directly affect the inflation rate, which in turn increases the uncertainty of the money multiplier and ultimately affects China's money supply mechanism and the effect of monetary policy transmission. Xie Xing (2019) [14] believes that the Central bank digital currencies will have a substitution or crowding-out effect on cash in circulation as well as demand bank deposits in the initial period of issuance, leading to a decline in the quantity of base money and an increase in the number of deposit reserves in the short term, and the money multiplier will become larger. Yang Dong et al. (2020)[16]believe that the issuance of the Central bank digital currencies will make it more difficult to predict the velocity of currency circulation and interest rates, and the effectiveness of China's monetary policy will face challenges.

In summary, scholars on the Central bank digital currencies from various aspects of the research, but on the impact of the Central bank digital currencies on monetary policy research is less and most of the research is still at the theoretical level, the lack of detailed, logical, and rigorous quantitative analysis of the argumentation process, and less literature combining theoretical and empirical analysis, the above shortcomings also reserved space for the research of this paper.

### 3 Theoretical analysis

1. the impact of legal tender on monetary policy
(1) the Central bank digital currencies enhances the effectiveness of China's monetary policy
First, the increased operational room for monetary policy will result from the creation of central bank digital currency. The issue and use of central bank digital currencies can lower the costs associated with keeping and moving money around, which helps to break the zero interest rate floor restriction and expand the monetary policy's operational space. Second, the issuance of the Central bank's digital currencies will improve the timeliness and accuracy of market economic information, and the traceability of the country's central bank's digital currencies will allow for real-time monitoring of the flow and transaction dynamics in currency circulation. In addition, the programmability of the Central bank digital currencies in technical realization gives the central bank the ability to directly regulate the Central bank digital currencies circulating in society, strengthening the precision of the central bank's digital currency placement in the industry and the control of the ultimate goal system of monetary policy. Thirdly, the issuance of China's the Central bank digital currencies has enriched the central bank's monetary operation tools, improved the central bank's ability to monitor the supply and demand of money, and prompted the central bank to regulate the legal reserve ratio as a monetary policy tool more accurately, which not only improves the transmission efficiency of monetary policy but also improves the precision and effectiveness of the central bank's economic regulation and control.

(2) Impact of the Central bank digital currencies on the Intermediation Objectives of Monetary Policy

In terms of money supply, firstly, China's Central bank digital currencies will intensify the uncertainty of narrow money multiplier and broad money multiplier, leading to the uncertainty impact on money supply. Secondly, the issuance of China's Central bank digital currencies will help to increase the amount of commercial banks' deposit reserves with the central bank, which will help to enhance the central bank's control over the money supply.

In terms of interest rates, the issuance of the Central bank digital currencies will accelerate the speed of conversion between financial assets and reduce the cost of conversion, and the deposit interest rates between major banks will gradually converge to an equilibrium state, which will enable the interest rate level to more truly reflect the relationship between supply and demand in the money market, and will be conducive to causing adjustments to the structure of the assets of the microeconomic entities through changes in interest rates. At the same time, with the steady progress of China's interest rate marketization, the issuance of Central bank digital currencies is conducive to the central bank's timely mastery of the application of LPR for new loans issued by banking institutions, improving the timeliness of the impact of LPR on credit interest rates, and enabling the central bank to influence the level of credit interest rates to regulate the economy. The issuance of Central bank digital currencies can enhance the ability of interest rates to reflect the supply and demand of funds in the financial market, improve the efficiency and timeliness of the adjustment of the benchmark interest rate, and contribute to a more efficient interest rate transmission mechanism in China.

**2. the impact of legal tender on the money multiplier**

In this paper, assuming that r is the reserve for deposits, the cash leakage rate is k, and the ratio of time and demand deposits is t, the narrow money multiplier and the broad money multiplier are, respectively:

$$m_1 = \frac{1+k}{r+rt+k} \tag{1}$$

$$m_2 = \frac{1+k+t}{r+rt+k} \tag{2}$$

(1) Impact of legal tender on cash leakage rate

First, Central bank digital currencies has the characteristics of high security, convenient payment, etc. Its issuance and circulation will replace a part of the cash in circulation in the short term, so that the cash in circulation M0 will be reduced, i.e., the cash leakage rate k will be reduced. Secondly, with the increasing income of the public, the marginal propensity to consume shows a decreasing law, i.e., the proportion of cash in the structure of public monetary assets will decrease. According to equations (1) and (2) to find the partial derivation of the cash leakage rate k. Since the fixed activity ratio t is a non-negative number without upper bound, equation (3) can be positive or negative.

$$\frac{\partial m_1}{\partial k} = \frac{r(1+t)-1}{[r+rt+k]^2} \tag{3}$$

$$\frac{\partial m_2}{\partial k} = \frac{(r-1)(1+t)}{[r+rt+k]^2} < 0 \tag{4}$$

(2) Impact of legal tender on reserve requirement ratio

First, the issuance of legal tender in China can improve the payment and settlement system in the interbank market, enhance the ability of commercial banks to allocate assets, increase the activity of the interbank market, improve the efficiency of transactions, and improve the ability of commercial banks to withstand liquidity risks. Second, Central bank digital currencies will increase the opportunity cost of holding excess reserves for commercial banks, prompting them to reduce their excess reserves. Both of these effects will cause commercial banks to reduce the excess reserve ratio. According to equations (1) and (2) to find the partial derivation of the deposit reserve ratio r.

$$\frac{\partial m_1}{\partial r} = -\frac{(1+t)(1+r)}{[r+rt+k]^2} < 0 \tag{5}$$

$$\frac{\partial m_2}{\partial r} = -\frac{(1+k+t)(1+t)}{[r+rt+k]^2} < 0 \tag{6}$$

(3) Impact of legal tender on the fixed-to-live ratio

The conversion between non-interest-bearing Central bank digital currencies and other interest-bearing assets is not significant, so the impact of Central bank digital currencies on the number of bank time deposits is small, and non-interest-bearing Central bank digital currencies does not have a significant crowding-out effect on interest-paying time deposits. At the early stage of issuance, the substitution effect of Central bank digital currencies for cash in circulation dominates, so its crowding-out effect on demand deposits is not yet apparent. Therefore, the impact of Central bank digital currencies on the fixed-to-demand ratio is uncertain at the initial stage of issuance. With the continuous promotion and popularization of Central bank digital currencies,the increasing substitution effect on cash, the overall increase in its recognition by the public, and the continuous introduction of new digital financial products by banks.

Then it will have a larger crowding effect on the demand deposits of commercial banks, so the number of demand deposits will be relatively reduced, and the fixed-to-live ratio will be reduced. According to equations (1) and (2), the fixed-to-live ratio t is derived.

$$\frac{\partial m_1}{\partial t} = -\frac{r(1+k)}{[r+rt+k]^2} < 0 \qquad (7)$$

$$\frac{\partial m_2}{\partial t} = \frac{k(1-r)}{[r+rt+k]^2} > 0 \qquad (8)$$

(4) Impact of legal tender digital currency on money multiplier

This paper estimates the change in money multiplier with the help of the differential principle. Assuming that the changes in k, r, and t are Δk, Δr, and Δt, respectively, and when Δk, Δr, and Δt tend to infinity, it can be assumed that Δk=dk, Δr=dc, and Δt=dt, then it can be obtained:

$$\Delta m_1 = \frac{[r(1+t)-1]\Delta k - [(1+t)(1+r)]\Delta r - r(1+k)\Delta t}{[r+rt+k]^2} \qquad (9)$$

$$\Delta m_2 = \frac{[(r-1)(1+t)]\Delta k - [(1+k+t)(1+t)]\Delta r + k(1-r)\Delta t}{[r+rt+k]^2} \qquad (10)$$

The above theoretical analysis leads to the conclusion that the issuance of Central bank digital currencies in China will have a positive impact on the narrow money multiplier and the broad money multiplier in general.

## 4 Model construction and data selection

**1. Model construction**

This paper combines the Central bank digital currencies, money multiplier, cash leakage rate, fixed-to-live ratio excess reserve ratio, and other indicators to construct the following empirical model to explore the impact of Central bank digital currencies on the narrow money multiplier and broad money multiplier.

$$m = \alpha_0 + \alpha_1 MDI + \alpha_2 k + \alpha_3 t + \alpha_4 e + \varepsilon \qquad (11)$$

Where m refers to the narrow money multiplier and the broad money multiplier; MDI stands for the issuance size of Central bank digital currencies; k stands for the cash leakage rate; the ratio of time deposits to demand deposits is denoted by t; and e stands for the excess reserve ratio of commercial banks.

**2. Selection of variables**

(1) Explained variable money multiplier m: narrow money multiplier m1 and broad money multiplier m2, which are obtained by the ratio of narrow money supply M1 and broad money supply M2 to the base currency B, respectively. The data are obtained from the official website of the People's Bank of China.

(2) Core Explanatory Variable Central bank digital currencies: as the digital RMB is still in the pilot stage, it has not been formally launched to the public, and the specific issuance data is unknown. At present, China's Central bank digital currencies is mainly a substitute for the transfer of M0 and third-party mobile payments, and Central bank digital currencies is currently mainly traded through the binding of bank cards, according to Shi Xinlu et al. (2018)[10]. The measurement method and data processing of third-party payment as well as electronic money, comprehensive consideration of China's future issuance of digital RMB to third-party payment and credit cash is a gradual replacement. So this paper constructs the MDI instead of the future issuance of Central bank digital currencies in the future, where MDI = (online payment amount + bank card transfer amount + bank card consumption balance) / GDP, the size of the MDI to a certain extent also represents the breadth of the Central bank digital currencies issuance to be used in the future. Source of data: The People's Bank of China (PBOC) quarterly "Payment System Operation General Situation" and Wind database.

(3) Control variables: this paper selects three control variables: first, the cash leakage rate k = cash in circulation/bank deposits, M0 data source from the People's Bank of China website, bank deposits data source from the Wind database; second, the fixed-to-live ratio t = (M2-M1)/(M1-M0), data source from the People's Bank of China website, "Depository Firms at a Glance"; and third, the excess deposit reserve ratio e, data source from the Wind database.

Table 1 Variable definitions and descriptive statistics

| Variable Symbol | Variable Definition | Mean | Maximum | Minimum | Standard Deviation |
|---|---|---|---|---|---|
| $m_1$ | Narrow money multiplier | 1.5600 | 1.9661 | 1.1626 | 0.2839 |
| $m_2$ | Broad money multiplier | 5.3191 | 7.8654 | 3.7771 | 1.2790 |
| MDI | Central bank digital currencies size | 7.7093 | 11.2676 | 3.5044 | 2.7615 |
| k | Cash leakage rate | 0.1882 | 0.0268 | 0.0090 | 0.0036 |
| t | Fixed Livelihood Ratio | 2.8409 | 3.6114 | 2.8483 | 0.3715 |
| e | Excess Reserve Ratio | 0.0189 | 0.0268 | 0.0090 | 0.0036 |

Source: Wind database and People's Bank of China.

# 5 Analysis of empirical results

### 1. Smoothness test of variables

Because the observations of time series often have cyclic cyclic changes, in this paper, we firstly seasonalize m1, m2, MDI, k, t, e. Through the method of CensusX-12 seasonal adjustment in Eviews10 software, we obtain m1_sa, m2_sa, MDI_sa, k_sa, t_sa, e_sa, respectively. Smoothness test. In this paper, ADF method is used to conduct unit root test for m1_sa, m2_sa, MDI_sa, k_sa, t_sa, and e_sa, and as shown in Table 2, all the variables are smooth at the 5% level, and all of them are first-order single-integrated.

Table 2 Smoothness test for each variable

| Variables | Test form (C, T, L) | t-value | P-value | Test result |
|---|---|---|---|---|
| $m_1\_sa$ | (C,0,5) | -1.3196 | 0.6127 | non-stationary |
| $Dm_1\_sa$ | (0,0,4) | -2.0390 | 0.0409 | steady |
| $m_2\_sa$ | (C,T,0) | -2.2262 | 0.4636 | non-stationary |
| $Dm_2\_sa$ | (0,0,0) | -3.1322 | 0.0025 | steady |
| MDI_sa | (C,0,0) | -1.5433 | 0.5039 | non-stationary |
| DMDI_sa | (0,0,0) | -5.8373 | 0.0000 | steady |
| k_sa | (C,0,8) | -1.9837 | 0.2926 | non-stationary |
| Dk_sa | (0,0,2) | -1.9951 | 0.0450 | steady |
| t_sa | (C,T,5) | -2.6397 | 0.2656 | non-stationary |
| Dt_sa | (C,0,0) | -3.4720 | 0.0129 | steady |
| e_sa | (C,0,1) | -2.5569 | 0.1087 | non-stationary |
| De_sa | (0,0,0) | -9.7198 | 0.0000 | steady |

### 2. Johansen cointegration test

The optimal lag order under the VAR model is first determined, as shown in Table 3, the lag order of both the narrow money multiplier variable group and the broad money multiplier variable group is order 3. In this paper, Johansen test is performed on the original time series of the variables based on VAR (3).

Table 3 Results of optimal lag order test

| Lag | LogL | LR | FPE | AIC | SC | HQ |
|---|---|---|---|---|---|---|
| Narrow money multiplier variable set ($m_1$_sa、MDI_sa、k_sa、t_sa、e_sa) | | | | | | |
| 0 | 298.7479 | NA | 3.33e-12 | -12.2395 | -12.0446 | -12.1658 |
| 1 | 575.8481 | 484.9254 | 9.18e-17 | -22.7437 | -21.5742* | -22.3017 |
| 2 | 615.2663 | 60.7698 | 5.22e-17 | -23.3443 | -21.2003 | -22.5342* |
| 3 | 647.9170 | 43.5343* | 4.15e-17* | -23.6632* | -20.5445 | -22.4847 |
| 4 | 667.3748 | 21.8899 | 6.29e-17 | -23.4323 | -19.3390 | -21.8854 |
| Variable group of broad money multipliers ($m_2$_sa、MDI_sa、k_sa、t_sa、e_sa) | | | | | | |
| 0 | 230.9715 | NA | 5.60e-11 | -9.4155 | -9.2206 | -9.3418 |
| 1 | 515.244 | 497.4769 | 1.15e-15 | -20.2185 | -19.0490* | -19.77654 |
| 2 | 554.9673 | 61.2401 | 6.43e-16 | -20.8319 | -18.6879 | -20.0217* |
| 3 | 586.6752 | 42.2772* | 5.33e-16* | -21.1115* | -17.9928 | -19.9329 |
| 4 | 604.4395 | 19.9848 | 8.66e-16 | -20.8099 | -16.7167 | -19.2631 |

Table 4 Results of Johansen's cointegration test

| original hypothesis | Characteristic Roots | Trace statistic | 0.05 Critical Value | Maximum eigenroot statistic (p-value) |
|---|---|---|---|---|
| Narrow money multiplier variable set ($m_1$_sa、MDI_sa、k_sa、t_sa、e_sa) | | | | |
| None* | 0.7136 | 145.6395 | 79.3414 | 0.0000 |
| At most 1* | 0.5309 | 84.3651 | 55.2458 | 0.0000 |
| At most 2* | 0.4099 | 47.2785 | 35.0109 | 0.0016 |
| At most 3* | 0.3078 | 21.4279 | 18.3977 | 0.0183 |
| At most 4 | 0.0671 | 3.4034 | 3.8415 | 0.0651 |
| Variable group of broad money multipliers ($m_2$_sa、MDI_sa、k_sa、t_sa、e_sa) | | | | |
| None * | 0.695415 | 138.8449 | 69.81889 | 0.0000 |
| At most 1 * | 0.578983 | 80.59343 | 47.85613 | 0.0000 |
| At most 2 * | 0.37076 | 38.20435 | 29.79707 | 0.0043 |
| At most 3 * | 0.229555 | 15.5055 | 15.49471 | 0.0498 |
| At most 4 | 0.054132 | 2.726973 | 3.841465 | 0.0987 |

At a 5% significance level, it shows that there is a stable equilibrium relationship between the narrow money multiplier variable group and the broad money multiplier variable group in the long run. And the increase of legal tender will positively affect the money multiplier. The long-term cointegration relationship is:

$$m_1 = 0.3163 MDI + 25.6739k - 0.8104t - 5.6796e \qquad (12)$$

$$m_2 = 4.4056 MDI + 322.4528k - 5.5866t - 3.2978e \qquad (13)$$

### 3. VECM model

Vector Error Correction Models (VECM) are developed to analyze the short-run cointegration relationship between the group of narrow money multiplier variables and the group of broad money multiplier variables. The first group is chosen as the main cointegration equation in both this paper.

For the narrow money multiplier variable group, the short-term error correction term is:

$$ECM_{1,t-1} = m_{1,t-1} + 0.5476 MDI_{t-1} + 36.2938 k_{t-1} - 0.7696 t_{t-1} + 1.6089 e_{t-1} - 7.4117 \quad (14)$$

For the set of broad money multiplier variables, the short-run error correction term is:

$$ECM_{2,t-1} = m_{2,t-1} + 4.4056 MDI_{t-1} + 322.4528 k_{t-1} - 5.5866 t_{t-1} - 3.2978 e_{t-1} + 73.9386 \quad (15)$$

### 4. AR root test

In order to verify whether the constructed VECM model system is stable, the eigenvalues are tested. The results are shown in Figure 1, indicating that the VECM model constructed in this paper is stable.

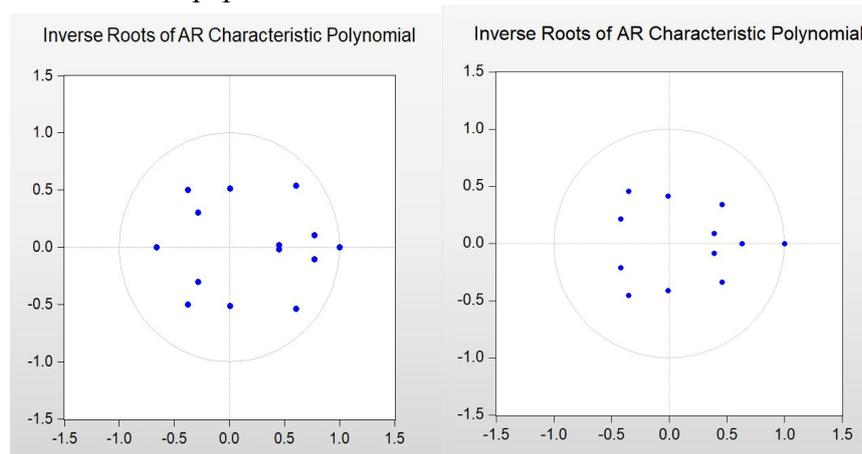

Figure 1 Roots of identity test for the group of narrow money multiplier variables (left) and the group of broad money multiplier variables (right)

### 5. Impulse response analysis

Impulse response analysis is conducted for the two sets of VECM models with a lag of 20 periods, and the figure shows the impulse response of Central bank digital currencies on the money multiplier. When m1 and m2 are subjected to one positive shock of MDI, it indicates that legal digital money has a positive impact on the narrow money multiplier and the broad money multiplier.

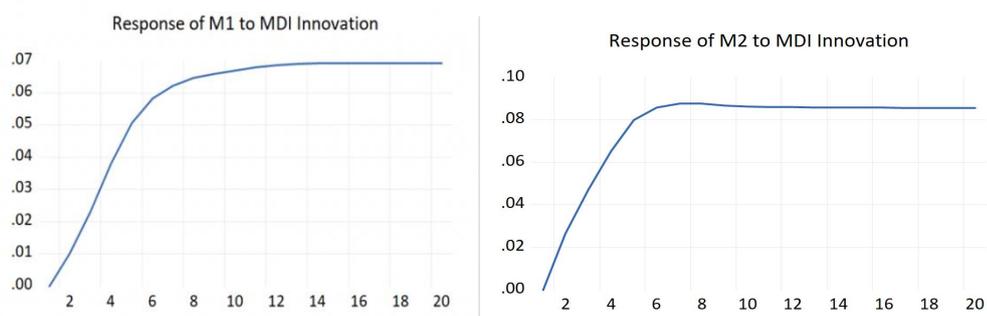

Figure 2 Narrow money multiplier impulse response (left) Broad money multiplier impulse response (right)

### 6.Variance decomposition

Table 5 Variance decomposition of the narrow money multiplier

| Period | S.E. | $m_1$ | MDI | k | t | e |
|---|---|---|---|---|---|---|
| 1 | 0.02849 | 100 | 0 | 0 | 0 | 0 |
| 2 | 0.044426 | 91.63703 | 4.945053 | 2.780687 | 0.366727 | 0.270501 |
| 3 | 0.063106 | 79.20966 | 15.94856 | 4.09529 | 0.557399 | 0.189086 |
| 4 | 0.088345 | 68.30619 | 27.39343 | 3.878279 | 0.324834 | 0.097275 |
| 5 | 0.116932 | 60.34447 | 36.03558 | 3.362906 | 0.187656 | 0.069381 |
| 6 | 0.145027 | 55.30734 | 41.60145 | 2.849733 | 0.163578 | 0.077895 |
| 7 | 0.170997 | 52.09366 | 45.08094 | 2.510624 | 0.206291 | 0.108481 |
| 8 | 0.19458 | 49.77919 | 47.48545 | 2.321657 | 0.268856 | 0.144849 |
| 9 | 0.215918 | 48.02175 | 49.25661 | 2.21939 | 0.328708 | 0.173541 |
| 10 | 0.235607 | 46.64832 | 50.61509 | 2.162266 | 0.380264 | 0.194067 |

Table 6 Variance decomposition of broad money multiplier

| Period | S.E. | $m_2$ | MDI | k | t | e |
|---|---|---|---|---|---|---|
| 1 | 0.070803 | 100 | 0 | 0 | 0 | 0 |
| 2 | 0.120936 | 76.63151 | 4.351097 | 9.143479 | 6.417898 | 3.456018 |
| 3 | 0.172291 | 58.23846 | 6.306513 | 13.76561 | 13.04976 | 8.639655 |
| 4 | 0.230284 | 50.78267 | 6.089468 | 14.48311 | 17.94368 | 10.70108 |
| 5 | 0.289944 | 46.83841 | 5.326087 | 14.89937 | 21.38247 | 11.55367 |
| 6 | 0.34507 | 44.44808 | 4.582843 | 15.37129 | 23.30627 | 12.29151 |
| 7 | 0.395051 | 43.00411 | 4.044219 | 15.76957 | 24.30106 | 12.88104 |
| 8 | 0.440012 | 41.96406 | 3.70746 | 16.09877 | 24.84975 | 13.37996 |
| 9 | 0.480519 | 41.14814 | 3.508982 | 16.36275 | 25.16829 | 13.81183 |
| 10 | 0.517605 | 40.50616 | 3.393471 | 16.55909 | 25.3774 | 14.16388 |

## 6 Conclusion and Recommendations

This paper analyzes the impact of Central bank digital currencies on China's monetary policy and the various variables of the money multiplier through theoretical analysis, and based on the theoretical analysis of the quarterly time series data selected from 2010-2022, the establishment of the VECM model for empirical analysis. The empirical results show that the Central bank digital currencies has a certain impact on the cash leakage rate, the excess reserve ratio and the fixed-to-live ratio, and the changes in the fixed-to-live ratio are more complicated and difficult to determine, which will increase the difficulty of the central bank in formulating monetary policy. Based on the above conclusions, relevant suggestions are put forward according to the development status of Central bank digital currencies in China:

First, the central bank should continue to strengthen the underlying technical support foundation of the Central bank digital currencies. First, it is necessary to strengthen the research and development and innovation of blockchain technology in the research and development and management of China's Central bank digital currencies, improve the digital currency technical standard system, continuously optimize the design of the digital renminbi system, maintain the overall technological sophistication, and provide a solid bottom-up support for the innovation of the Central bank digital currencies business and application, and then guarantee that the digital currency system can stably realize the characteristics of the Central bank digital currencies, and ensure that it can be safe and stable circulation; continue to improve the security management of its operation system, and provide a perfect technology and service system for the popularization of Central bank digital currencies. Second, accelerating the construction of a digital central bank will help improve the timeliness and precision of macroeconomic policies and realize real-time sharing of financial market information.

Secondly, the construction of application scenarios should be strengthened, and the popularization of Central bank digital currencies should be actively promoted. Although China's Central bank digital currencies is accelerating its entry into the homes of ordinary people, the public's awareness and acceptance of Central bank digital currencies as a novelty needs to be improved. In terms of application scenarios, it is necessary to gradually form a full range of application scenarios for Central bank digital currencies in daily consumption, payment, transportation, education and medical care, culture, sports and tourism. In the promotion process, emphasis has been placed on simplifying the payment process, realizing diversified payment methods for different groups of people, and including the elderly population in the scope of application of Central bank digital currencies to narrow the digital divide.

Fourth, strengthen the legislative framework and policies supporting the industry in relation to Central bank digital currencies. To control the development of Central bank digital currencies, provide safeguards for their long-term benign development from the system, consolidate the rule of law and regulatory basis for their development in China, supplement China's Regulations on the Administration of the Renminbi, and expand the contents of Central bank digital currencies, it is necessary for the Central Bank and relevant departments to develop laws and regulations.In order to create a smooth and sensitive monetary policy interest rate transmission mechanism and ultimately achieve the monetary policy objectives through the regulation of market interest, the central bank should enhance the market benchmark interest rate system with Shibor as the core. It should also gradually switch the monetary policy control mode to one that is interest rate-oriented. The central bank can take use of the novelty of monetary policy instruments provided by the advent of Central bank digital currencies to better fulfill the purpose of monetary policy.

  Fourth, strengthen the legislative framework and policies supporting the industry in relation to Central bank digital currencies. To control the development of Central bank digital currencies, provide safeguards for their long-term benign development from the system, consolidate the rule of law and regulatory basis for their development in China, supplement China's Regulations on the Administration of the Renminbi, and expand the contents of Central bank digital currencies, it is necessary for the Central Bank and relevant departments to develop laws and regulations.

[References]

[1] Barrdear&Kumhof. The Macroeconomics of Central Bank Issued Digital Currencies. Staff Working Paper, Bank of England, 2016.

[2] Bank for International Settlements. "Central bank digital currencies, BIS Committee on Payments and Market Infrastructures". Market Committee

[3] Bordo MD and Levin AT. Central Bank Digital Currency and the Future of Monetary Policy. NBER Working Paper No.23711, 2017.

[4] Blakstad S and Robert A. Fin Tech Revolution. Cham, Switzerland Springer, 2018.

[5] Du Yongshan,Gao Jie. The impact of central bank digital currency issuance on China's monetary policy framework[J]. Enterprise Economy,2022,41(04):65-75.

[6] Fang Xiancang,Huang Siyu. Digital currency and the transformation of China's monetary policy[J]. Academic Forum,2020,43(02):91-101.

[7] Fan Yifei. Theoretical basis and architectural choice of legal digital currency in China[J]. China Finance,2016(17):10-12.

[8] Koning J.P.,2017.Evolution in cash and payments:comparing old and new ways of designing central bank payments systems,cross-border payments networks,and remittances.R3 Reports.

[9] Lin Chun. Legal tender digital currency issuance and the effectiveness of China's monetary policy[J]. Journal of Shenzhen University (Humanities and Social Sciences Edition),2019,36(05):77-86.

[10] SHI Xinlu,ZHOU Zhengning. Electronic payments, money substitution and money supply[J]. Research in Financial Economics,2018,33(04):24-34.

[11] Stiglitz J. E. Macro-Economic Management in an Electronic Credit/Financial System. NBER Working Paper, 2017.

[12] Wu Xinhong,Pei Ping. Legal Digital Currency: Theoretical Foundations, Operational Mechanisms and Policy Effects[J]. Journal of Soochow University (Philosophy and Social Science Edition),2022,43(02):104-114.

[13] Wu ZF. Blockchain and digital currency issuance[J]. International Finance,2016(09):20-23.

[14] Xie Xing,Feng Sixian. Theoretical research on the impact of legal digital currency on China's monetary policy[J]. Economist,2019(09):54-63.

[15] Yao Qian. China's legal digital currency prototype conception[J]. China Finance,2016(17):13-15.

[16] YANG Dong,CHEN Zheli. Research on the positioning and nature of legal digital currency[J]. Journal of Renmin University of China,2020,34(03):108-121.